\documentclass[pra,superscriptaddress,twocolumn,superscriptaddress,amsmath,noshowpacs]{revtex4}

\usepackage{epsf,latexsym,bbm,bm}
\usepackage{amssymb,amsmath}
\usepackage{times,graphics}
\usepackage{color,cancel}
\usepackage{epsfig,ulem}
\usepackage{upgreek}
\usepackage{yfonts}[1998/10/03]
\usepackage[T1]{fontenc}
\usepackage{booktabs,appendix}
\newcommand{\bra}[1]{\langle #1 |}
\newcommand{\ket}[1]{| #1 \rangle}

\def\6{\langle}
\def\9{\rangle}

\newcommand\bA{{\bf{A}}}
\newcommand\bB{{\bf{B}}}
\newcommand\bE{{\bf{E}}}
\newcommand\bF{{\mathbf{F}}}
\newcommand\bG{{\bf{G}}}
\newcommand\bJ{{\bf{J}}}
\newcommand\bK{{\bf{K}}}
\newcommand\bN{{\bf{N}}}
\newcommand\bL{{\bf{L}}}

\newcommand\bS{{\bf{S}}}
\newcommand\bX{{\bf{X}}}

\newcommand\boa{{\bf{a}}}

\newcommand\bg{{\bf{g}}}
\newcommand\bk{{\mathbf{k}}}
\newcommand\bn{{\mathbf{n}}}
\newcommand\bp{{\bf{p}}}
\newcommand\bs{{\bf{s}}}

\newcommand\bv{{\mathbf{v}}}
\newcommand\bw{{\mathbf{w}}}
\newcommand\bu{{\mathbf{u}}}
\newcommand\bx{{\mathbf{x}}}
\newcommand\bq{{\mathbf{q}}}

\newcommand\sS{{\mathsf{S}}}
\newcommand\sW{{\mathsf{W}}}

\newcommand\cO{{\cal{{O}}}}
\newcommand\cL{{\cal{{L}}}}
\newcommand\cS{{\cal{{S}}}}

\newcommand\cW{{\cal{{W}}}}

\def\etal{\textit{et al.}}

\newcommand\balp{{\boldsymbol{\upalpha}}}
\newcommand\bsig{{\boldsymbol{\upsigma}}}

\newcommand\bmsig{{\bm\Sigma}}

\def\half{{\tfrac{1}{2}}}

\newcommand\hb{{\hat{b}}}
\newcommand\hd{{\hat{d}}}

\def\hH{{\hat{H}}}
\def\hS{{\hat{S}}}
\def\hT{{\hat{T}}}

\def\hPi{{\hat{\Pi}}}

\def\hpsi{{\hat{\psi}}}

\def\hbS{{\bf\hat{S}}}

\newcommand{\pad}{\partial}
\newcommand{\ra}{{\rightarrow}}
\newcommand\bnab{{\mathbf{\nabla}}}

\newcommand{\be}{\begin{equation}}
\newcommand{\ee}{\end{equation}}
\newcommand{\ba}{\begin{eqnarray}}
\newcommand{\ea}{\end{eqnarray}}
\newcommand{\tr}{\mathrm{tr}}

\begin{document}
\title{Spin and localization of relativistic fermions and uncertainty relations}

\author{Lucas C. C\'{e}leri}
\email{lucas@chibebe.org}
\author{Vasilis Kiosses}
\email{kiosses.vas@gmail.com}
\affiliation{Instituto de F\'{\i}sica, Universidade Federal de Goi\'{a}s, Goi\^{a}nia, GO, Brazil}

\author{Daniel R. Terno}
\email{daniel.terno@mq.edu.au}
\affiliation{Department of Physics \& Astronomy, Macquarie University, NSW 2109, Australia}

\begin{abstract}
We discuss  relations between several  relativistic spin observables and derive a Lorentz-invariant characteristic of a reduced spin density matrix. A relativistic position operator that satisfies all the properties of its non-relativistic analog does not exist. Instead we propose two causality-preserving positive operator-valued measures (POVM) that are based on projections onto one-particle and antiparticle spaces, and on the normalized energy density. They predict identical expectation values for  position.  The variances differ by less than a quarter of the squared de Broglie wavelength and coincide in the nonrelativistic limit. Since the resulting statistical moment operators are not  canonical conjugates of momentum, the Heisenberg uncertainty relations need not hold. Indeed, the energy density POVM leads to a lower uncertainty.  We reformulate the standard equations of the spin dynamics by explicitly considering the  charge-independent acceleration,  allowing    a  consistent treatment of backreaction and  inclusion of a weak gravitational field.
\end{abstract}
\maketitle

\section{Introduction}


{Spin, position and momentum --- of individual particles and their aggregates --- are both dynamical variables and observables 
in atomic, nuclear and condensed matter physics. The nonrelativistic spin $\hbar\bsig/2$ is  a textbook embodiment of quantum formalism \cite{peres:95} while its eigenstates are the prototype  of a quantum bit \cite{qinfo}.}  However, interactions of the standard model
are described by   couplings of fields. While the fields form representations of the Lorentz group and carry spin labels, neither spin nor momentum are dynamical variables and their role as observables is established through additional considerations \cite{haag, ll4, bogo:90}.

  Construction of relativistic position  is fraught with technical difficulties and compromises between different reasonable  requirements (see, e.g., \cite{haag,pryce:48,nw:49,heg:85,bu:99,bk:03,pt:04,t:14} and references therein). There is no unique way to describe localization of a relativistic particle, even when particles are unambiguously defined. There is no unique spin operator as well. Indeed, there are at least seven \cite{7op}.

There are several reasons for proliferation of spin and position variables. Conceptually, these are emergent objects that are fleshed out in the descend from quantum field theory through relativistic and nonrelativistic quantum mechanics to the level of classical physics. Pragmatically, decomposition of the total angular momentum $\bJ=\bL+\bS$ into the orbital angular momentum $\bL$ and the spin $\bS$ parts,
 \be
 \bJ=\bx \,\times\,\bp+\bS, \label{sx}
 \ee
with $\bp$ being the momentum, ensures that each alternative proposal for spin results in a corresponding  position $\bx$, and vice versa \cite{pryce:48}.

The acceptable level of approximation in localization and spin estimation is determined by the actual experimental set-up. With the typical (relative) energy spreads of the order of $10^{-3}-10^{-5}$ \cite{pdata} and the intrinsic spatial resolution on the scale between micrometer and millimeter \cite{pdet}, the motion of particles in accelerators can be described classically, and various quantum effects treated as perturbations, albeit the ones that may be critically important for the actual functioning of the machine \cite{accel1,accel2,st:86}. For the purposes of beam manipulation and scattering analysis the spin states are conveniently characterized by helicity, rest frame spin or four-polarization vector \cite{polrev}, and the spin is evolved under the assumption of a given particle trajectory \cite{accel1,accel2}

Dynamics of mixed quantum-classical systems, while often a convenient  approximation, leads to inconsistencies (\cite{qclad} and references therein). Order by order calculations of spin backreaction on the trajectory improve precision, but bring in higher-order derivatives, spurious solutions and difficulty in formulating Hamiltonian dynamics \cite{psk:00}.

Several research directions motivate the renewed interest in relativistic spin and localization. Bell-type inequalities and their experimental violations are arguably one of the most important results in quantum foundations \cite{peres:95}. However, it was shown that these violations critically depend on the type of spin operator  {involved \cite{cz:97, crw:09} even when the finite wave-packet width effects \cite{ga:03,crpsw:14}} are not taken into account, since not all such operators \cite{t:03} satisfy the necessary commutation relations \cite{suwe}. In general,  properties of relativistic spin are responsible for many features of relativistic quantum information theory that distinguish it from its nonrelativistic counterpart \cite{pt:04,sv:12n}.

{There are proposals to separate charged particles of different polarizations in accelerator beams through the spin interaction with external fields in a storage ring  \cite{accel2, h:96}. Even considered as a purely theoretical exercise \cite{accel1}, these involve a subtle interplay between  continuous and discrete degrees of freedom.}

Ultra-high power lasers  producing tailored ultra-short pulses allow precise tracking of single relativistic electrons \cite{light-mat}. Spin and orbital angular momenta of electron beams and laser pulses produce spin-dependent probability distributions \cite{el}, while shaping of electron beams results in qualitatively new patterns of Cherenkov radiation \cite{kaminer:16}. Searches for spin-gravity coupling are part of the precision tests of gravity and  aim to discover the limitations and extensions of the standard model, general relativity and quantum gravity \cite{sme}.

{
In this paper we treat  three interlocked problems: descriptions of spin, position, and their evolution for relativistic fermions.} 
In Section~II we present the relationships between the rest frame  (Wigner, also referred to as Pryce \cite{7op} or Newton-Wigner \cite{crw:13}) spin, Dirac spin, Pauli-Lubanski vector and four-polarization. Most of these relations are either textbook results or their direct corollaries. However, together they provide a useful conceptual outlook as well as  computational tools that are applied in the following.  { A  survey of spin operators  can be found, e.g., in \cite{st:86, tha:92} and especially in \cite{bg:14} and \cite{7op}, that  include  exhaustive lists of references (in addition Ref.~\cite{bg:14} contains a historic sketch and \cite{7op} provides a comparative table of seven operators).}

Section~III analyses several proposals for localisation of Dirac fermions. Instead of constructing a {self-adjoint} position operator {the spectral decomposition of which is used to calculate probabilities of the measurement outcomes}, we focus on  positive operator-valued measures (POVMs). These are the most general mathematical structures that describe quantum measurements \cite{peres:95,qinfo}.
{We construct POVMs that are defined on one-particle space and result in causality-respecting probability distributions}. {In particular, we find that}  a localization scheme that is based on energy density leads to violation of the Heisenberg uncertainty relation.

 In Section~IV we reformulate the standard equations of spin dynamics by explicitly considering the effects  that are independent of the electric charge, thus allowing a more consistent treatment of backreaction and inclusion of weak gravity.  A final discussion is then presented in Section~V.

{In Appendix~\ref{ap:lorentz} we summarize the notation and conventions  for  Lorentz transformations.  Conventions for the Dirac equation and properties of related position and spin operators are presented in Appendix~\ref{ap:dirac}. Table~\ref{tableop} summarizes  notation and  classification of the matching spin and positions operators. Conventions for the quantum  fields and states are summarized in Appendix~\ref{ap:pf}.}

Unless specified otherwise we set $\hbar=c=1$. We work in the Minkowski spacetime with the metric $\mathrm{diagonal}(1,-1,-1,-1)$. Three-dimensional vectors (including  vectors of operators) are set in boldface, such as $\bp$ or $\bsig$. The four-dimensional spin  vector is denoted by sans font, $\sS=(S^0,\bS)$. Quantum field operators and their composites (that act on the fermionic Fock space or its one-particle restriction) are indicated by carets, such as $\hb_{p\sigma}$ or $\hbS$. The Einstein summation convention is used, with the Greek letters labelling the space-time indices 0,\ldots, 3, and the Latin indices running over 1,2,3 (Appendix~\ref{conven}).

\vspace{-2mm}
\section{Spin operators and density matrices}
\vspace{-2mm}
We begin from a survey of several popular quantities that are referred to as ``relativistic spin'' \cite{ll4,7op,polrev,st:86,tha:92,bg:14}. After discussing the relations between  spin variables/operators and the 4-vector of spin for  states of well-defined momentum, we  discuss the spin $4\times4$ density matrix.

In addition to different fonts that distinguish the four- and three-dimensional versions of spin-related quantities, different definitions of spin (Wigner, Dirac, Czachor, etc.), are indicated by the corresponding subscripts. Unless it leads to confusion we do not notationally distinguish classical vectors and vectors of the expectation values.
\begin{table}[tbp]
   \centering
    \caption{Brief summary of the matching spin and position operators. The type of the operators  according to Pryce \cite{pryce:48} is denoted as (x). The numbering according to Bauke \etal~\cite{7op} is indicated as X.}
    \label{tableop}
    \begin{tabular}{llc}
          \specialrule{0.2pt}{0pt}{1.4pt} \hline
     Spin~~~~~~~~~~~~~~~~~~~~~~~~~~~~ ~~~~  & Position~~~~~~~~~~~~~~~~~~~~~~~~~~ & Classification \\
     \hline
     $\bS_{\mathrm{D}}=\half \bmsig$~~ & ~~~~~$\bx$    &  A \\
      $\bS_{\mathrm{Cz}}$   & ~$\tilde\bx=\bq$     &  C\quad (c) \\
      $\bS_\mathrm{F}$  & ~~~~~$\bX$ & D\quad (d) \\
      $\bS_{\mathrm{W}}$ & ~~~~~$\tilde\bq$ & F\quad (e)\\

   \specialrule{0.2pt}{0pt}{1.3pt} \hline
   \end{tabular}

   \label{tab:booktabs}
   \vspace{-3mm}
\end{table}
\vspace{-2mm}
\subsection{Spin and polarization}

\vspace{-2mm}
Presentation of spin operators of massive particles is most conveniently couched in the semi-classical language, with particles having well-defined trajectories and thus  rest frames, and carrying spin, that also may be considered as a classical vector.   The (kinetic) momentum $p=(p^0,\bp)=mu=(1-v^2)^{-1/2} m(1,\bv)$ behaves as a classical parameter in the relevant spin transformations, hence the following analysis applies both to a classical particle with the momentum $p$ and to a momentum eigenstate $|p\9$. Unless stated otherwise, we consider free massive particles and fields of spin-$\half$. Localization and effective trajectories are discussed in Secs.~\ref{local3} and \ref{sec:dyn}.

Expectation value $\bs$ of a nonrelativistic spin  is obtained as
\be
\bs=\half\tr\rho\bsig, \qquad \rho=\half(I+\bn\cdot\bsig),
\ee
where $\rho$ is a $2\times 2$ spin density matrix, $\bsig$'s are the three Pauli matrices, $I$ is the identity,  and the Bloch vector $\bn=\tr\rho\bsig$ satisfies $0\leq|\bn|\leq 1$.

Assume that in the laboratory frame the particle has a four-momentum $p=(E_\bp,\bp)$, $E_\bp=p^0=\sqrt{\bp^2+m^2}$, and the transformation to the rest frame  is accomplished by the  standard boost $L_p^{-1}$ (see Appendix \ref{ap:lorentz}). The spin four-vector $\sS$ (also referred as four-polarization) is obtained by promoting the nonrelativistic spin $\bs$ 
to  a four-vector \cite{polrev,tamm:29} by setting
\be
\sS|_R=(0,\bs)|_R, 
\ee
 in the rest frame, and then in  any other reference frame by applying the corresponding  Lorentz transformation. In the laboratory frame {components of the four-vector spin} are given by
\be
S^\mu=L^\mu_p{}_\nu S|_R^\nu. \label{poldef}
\ee
The spin four-vector satisfies a number of useful identities, such as
\be
S^\mu p_\mu=0, \qquad \sS^2=-\bs^2, \qquad S^0=\frac{\bp\cdot\bS}{E_\bp}=\bv\cdot\bS. \label{polort}
\ee

From the group-theoretical point of view \cite{haag,bogo:90} elementary particles are distinguished by the values of two Casimir invariants (of the universal covering group) of the proper Poincar\'{e} group. These are the  mass $p^2=m^2$  and the square of the Pauli-Lubanski vector $\sW$,
\be
 W_\rho =\half\epsilon_{\lambda\mu\nu\rho} {p}^\lambda {M}^{\mu\nu}, \qquad \sW^2=-s(s+1),
\ee
where $s$ is an integer or a half-integer number, and the algebra generators are the four-momentum $p$   and the antisymmetric four-dimensional angular momentum $M^{\nu\lambda}=-M^{\lambda\nu}$. Using the generators of three-dimensional rotations and boosts,
\be
J^k:=\half\epsilon^{klm}M_{lm}, \qquad K^j:=M^{0j},
\ee
 respectively, the Pauli-Lubanski vector is given by
\be
 {W}^0= {\bp}\cdot{\mathbf{J}}, \qquad {\mathbf{W}}={p}^0{\mathbf{J}}+{\bp}\times{\bK}.
\ee

Two most widely used spin operators are the Wigner  and the Dirac-Pauli spin operators. The former is most conveniently introduced in the laboratory frame as \cite{bogo:90}
\be
  {\bS}_\mathrm{W}:=\frac{1}{m}\left( {\mathbf{W}}-\frac{ {W}^0\bp}{p^0+m}\right),
\ee
i.e. by taking the active view of the standard Lorentz boost $L_p$ as
\be
(0,\bS_{\mathrm{W}})=L_p^{-1}\cdot\sW/m.
\ee

On the other hand, seeing $L_p^{-1}$ as producing a coordinate transformation between the laboratory frame and the rest frame, we find that the Wigner spin is numerically equal to the rest frame spin,
\be
S^k_{\mathrm{W}}=s^k, \qquad k=1,2,3.
\ee
As a result, the Pauli-Lubanski vector is proportional to the spin four-vector,
\be
\sW=m\sS.
\ee

When the Lorentz transformation $\Lambda$ acts on the four-vector of spin, $\sS\ra\sS'=\Lambda\sS$, the Wigner spin is rotated,
\be
\bS_{\mathrm{W}}\ra\bS_{\mathrm{W}}'=R\bS_{\mathrm{W}},
\ee
where the three-dimensional  Wigner rotation $R$ is a nontrivial block of the Lorentz transformation
$\mathcal{W}:=L^{-1}_{\Lambda p}\Lambda L_p$. This is consistent with the transformation law for one-particle states  (see Appendix~\ref{ap:pf} for details).

{In the context of quantum field theory spin operators are expressed in terms of field operators (here those operators are denoted with hat).
The Wigner spin operator for free Dirac fermions is given by \cite{bogo:90}
\be
\hbS_{\mathrm{W}}=\half\sum_{\xi\zeta}\bsig_{\xi\zeta}\int\!d\mu(p)\big(\hb^\dag_{p \xi  }\hb_{p \zeta }+\hd^\dag_{p \xi  }\hd_{p \zeta}\big),
\ee
where $d\mu(p)=d^3\bp/(2\pi)^3(2p^0)$, and $\hb_{p\xi}$,\ldots, $\hd^\dag_{p\xi}$ are the annihilation and creation operators of particles and antiparticles, respectively (see Appendix~\ref{ap:pf} for conventions).

The Wigner spin satisfies the standard spin commutation relations
\be
[\hS^k_\mathrm{W},\hS^l_\mathrm{W}]=i\epsilon^{kl}_{~~m}\hS_\mathrm{W}^m.
\ee
The eigenvalues of $\hat{S}^3_\mathrm{W}$ are used to label one-particle states.  

On the space of solutions of the Dirac equation the Dirac spin \cite{ll4,pryce:48,tha:92,t:03,7op} is the simplest spin operator. In the standard representation it is just
\be
\half\bm{\Sigma}= \half\left(\begin{matrix}
\bsig & 0 \\
0 & \bsig \end{matrix}\right).
\ee
In terms of the fermion field $\hpsi(x)$  the Dirac spin is given by
\be
\hbS_{\mathrm{D}}=\half\!::\!\!\int\!d^3\bx\hpsi^\dag(x)\bm{\Sigma}\hpsi(x)\!::,
\ee
where $::$ denotes a normal ordering.

Consider a particle with a well-defined momentum and a particular spin,
\be
|\Psi\9=|p,\chi\9=\chi_1|p,+\half\9+\chi_2|p,-\half\9. \label{stsp}
\ee
Its associated Dirac spinor is
\be
u_\chi(\bp)=\chi_1 u^{1/2}_\bp +\chi_2 u^{-1/2}_ \bp. \label{spst}
\ee
Then  the expectation value of the Dirac spin operator $\hat{\bS}_\mathrm{D}$
\be
\bS_{\mathrm{D}}:=\frac{\6\Psi|\hbS_{\mathrm{D}}|\Psi\9}{\6\Psi|\Psi\9}=\frac{1}{4E_\bp}u^\dag_\chi(\bp){\bm\Sigma}u_\chi(\bp),
\ee
and the expectation value of the Wigner  spin is
\be
\bS_{\mathrm{W}}=\half\chi^\dag\bm\sigma\chi.
\ee

Using the properties of the spinors $u_\bp^\xi$ (Appendix \ref{spinor}) we obtain an explicit relationship between the two versions of spin,
\be
\bS_{\mathrm{D}}=\frac{m}{E_\bp}\bS_{\mathrm{W}}  +\frac{\bp(\bp\cdot\bS_{\mathrm{W}})}{E_\bp(E_\bp+m)}.
\ee
Once compared with Eq.~\eqref{poldef} we see that the Dirac spin is related to the spatial part of the four-polarization as
\be
\bS=\frac{E_\bp}{m}\bS_{\mathrm{D}}=\bS_{\mathrm{W}}+\frac{\bp(\bp\cdot\bS_{\mathrm{W}})}{m(E_\bp+m)}. \label{s2s}
\ee

 {The Wigner spin is a unique ``natural" relativistic extension of the nonrelativistic spin that is  linear in $\sW$ \cite{bogo:90,t:03}. The requirement of linearity also selects the Wigner spin out of four reasonable spin operators for a Dirac particle \cite{crw:13}.}

\subsection{Density matrices}
\vspace{-2mm}
For particle states with well-defined momentum the 4-vector of spin $\sS$ can be obtained from the $4\times 4$ polarization density matrix,
\be
S^\mu_p=\frac{1}{4m}\mathrm{tr}\big(\rho^{\mathrm{D}}_p\gamma^5\gamma^\mu\big), \label{traceS}
\ee
where $\rho^{\mathrm{D}}$ generalizes a pure state expression $u_\chi(\bp)\bar{u}_\chi(\bp)$ \cite{ll4,mw:55},
\be
\rho_p^{\mathrm{D}}=\sum_{\xi\zeta}c_{\xi\zeta}u^\xi_\bp\bar{u}^\zeta_\bp=\half(p\!\!/+m)(1-\half\gamma^5S\!\!\!/).
\ee
Here the coefficients $c_{\xi\zeta}$ satisfy the same conditions as the components of a usual $2\times 2$ spin density matrix, $\gamma^0,\ldots \gamma^5$,   are the Dirac $\gamma$ matrices,  $p\!\!/=\gamma^\mu p_\mu$, $\bar u=u^\dag\gamma^0$, $S\!\!\!/=\gamma^\mu S_\mu$, $\bar u=u^\dag\gamma^0$ (see Appendix \ref{spinor} for conventions).

Consider now a generic one-particle state
\be
|\Psi\9=\sum_{\xi=\pm\half}\int\!d\mu(p)f_\xi(\bp)|p,\xi\9, \label{genpsi}
\ee
where
\be
f_\xi(\bp)=\left(\begin{matrix}\chi_1(\bp)\\ \chi_2(\bp)\end{matrix}\right)f(\bp),
\ee
with $|\chi(\bp)|^2=|\chi_1(\bp)|^2+|\chi_2(\bp)|^2=1$ and $\int\! d\mu(p)|f(\bp)|^2=1$. The reduced $2\times 2$ spin density matrix, obtained by tracing out the momentum, reads
\be
\rho=\int d\mu(p)\chi(\bp)\chi^\dag(\bp)|f(\bp)|^2.
\ee

Due to dependence of the Wigner rotation $R$ on momentum, the reduced spin density matrix does not have a definite transformation law under Lorentz boosts. This is the basis of many results in relativistic quantum information theory about observer-dependence of spin entropy, distinguishability of spin states, and spin-spin entanglement \cite{pt:04}.

On the other hand,   the $4\times 4$ spinorial reduced density matrix
\be
\rho^{\mathrm{D}}:=\int d\mu(p)u_\chi(\bp)\bar{u}_\chi(\bp)|f(\bp)|^2,
\ee
still transforms as $\rho^\mathrm{D}\rightarrow B(\Lambda)\rho^\mathrm{D} B^{-1}(\Lambda)$, where $B(\Lambda)$ is the $(\half,\half)$ representation of the Lorentz group, $\Lambda\cdot u_\bp=B(\Lambda)u_{\bp}$. Hence the average four-vector of spin,
\begin{align}
\6S^\mu\9&=\frac{1}{4m}\int\!d\mu(p)\mathrm{tr}\big(u_\chi(\bp)\bar{u}_\chi(\bp)\gamma^5\gamma^\mu\big)|f(\bp)|^2\nonumber\\
&=\int\!d\mu(p)S^\mu_p |f(\bp)|^2.
\end{align}
is manifestly covariant, and $\6\sS\9^2$ is a Lorentz scalar that characterizes the state $|\Psi\9$.

However, this covariance cannot be exploited to make predictions of spin measurements any better. In any frame the relationship between the four-vector of spin and, e.g., Dirac spin can be used to calculate $\6\bS\9$,
\be
\6\bS\9=\int\!d\mu(p)\frac{E_\bp}{m}\bS_{\mathrm{D}}(\bp)|f(\bp)|^2\neq\6 \bS_{\mathrm{D}}\9\frac{\6\bE_\bp\9}{m},
\ee
and relationships \eqref{polort} will not generally hold. Since the actual interactions involve both position an momentum (see Section~IV), the knowledge of the covariant spin is not sufficient to make predictions for, e.g., the relativistic Stern-Gerlach experiment (compare with \cite{crpsw:14,sv:12a}).

\section{Position {POVM}}\label{local3}
\vspace{-1mm}
Taking fields  as fundamental and particles as emergent, it is not surprising to have a number of alternative methods to  localize them. {Moreover, analyzing dynamics of classical spinning particles and aiming to match their quantum mechanics with results of fundamental quantum theory results in additional crop of position variables \cite{psk:00,regge:74,dpm:14}}.

Our approach is motivated by the following. On the one hand, there are numerous obstacles for obtaining four (space-time) or three (space) self-adjoint position operators with the usual commutation relations. On the other hand, particle's position is not a dynamical variable in the field picture and thus does not have to be a part of a self-adjoint Hamiltonian. Hence we describe the localized detection events in terms of positive operator-valued measures (POVMs) \cite{t:14,bu:99}. A POVM constitutes a nonorthogonal decomposition of the identity by means of positive operators $\hPi(x)$, resulting in detection probabilities $P(x)=\tr \rho\hPi(x)$ for the set of events $\{x\}$ \cite{peres:95,qinfo, cov-povm}.

The resulting probability distributions cannot be localized too sharply: both localization in a bound region of space and fast-decaying exponential tails lead to violations of causality \cite{haag,heg:85,tha:92}.

The position operator of Newton and Wigner \cite{nw:49,haag,tha:92} [the (e) position operator of Pryce \cite{pryce:48}] and the associated single-particle probability density are known to lead to such violations. However, the standard Dirac probability density $\bar\psi\gamma^0\psi$ for positive or negative-energy solutions, or energy density do not \cite{bk:03}. Hence we study the POVMs that are built around these quantities. {Interesting features of localization of fermions in a cavity are described in \cite{klw:16}. Limitations of the localization POVM built from the field operators in general, and of the use of energy density in particular are discussed in \cite{pt:04,t:14}}.

Appendix~B 2 discusses some of the position operators for the Dirac equation, emphasizing the ones that are related to our POVMs. Field-theoretical constructions of the position operators for photons and the resulting uncertainty relations are discussed in \cite{bb:12}, while fermionic position operators that match the corresponding spin operators in the sense of Eq.~(1) were derived in \cite{crpsw:14,crw:13}. Our construction of the POVM follows the logic of \cite{t:14}.
\vspace{-1mm}
\subsection{Particle and antiparticle subspace POVM}
\vspace{-2mm}
Within the Dirac theory the standard multiplicative position operator $\bx$ mixes the spaces of positive and negative energy and thus is not observable \cite{pryce:48, tha:92}. A standard treatment is to separate it into the part that preserves the two subspaces and the part that connects them (Appendix~\ref{twoop}). Here we describe a field-theoretical analog of this procedure. Our goal is not a triple of operators $\hat{\bx}$, but a probability measure that allows us to calculate statistical moments  of a vector of classical random variables $\bx$.

Note that the operator
\be
\int\!d^3\bx\hpsi^{(+)\dag}(x)\hpsi^{(+)}(x)=I_\mathrm{1p},
\ee
acts as the identity on the one-particle subspace. Hence when restricted to the one-particle space (of particles and antiparticles), the  operator density
\be
\hPi_\bx(x):=\hpsi^{(+)\dag}(x)\hpsi^{(+)}(x)+\hpsi^{(-)}(x)\hpsi^{(-)\dag}(x)
\ee
is a positive decomposition of identity and thus a positive-operator valued measure \cite{peres:95,qinfo,cov-povm}. Since it is a local density, it is easy to see that it is an orthogonal decomposition of identity. Its expectation value on a generic one-particle state of Eq.~\eqref{genpsi} results in the standard Dirac probability density,
\be
\bra\Psi\hat\Pi_\bx(t,\bx)\ket\Psi=\big|\Psi(t,\bx)\big|^2,
\ee
where the four-component wave function $\Psi(x)$ that corresponds to the state $|\Psi\9$ of Eq.~\eqref{genpsi} is given by
\begin{align}
\Psi(t,\bx)=&\sum_\xi\!\int\!d\mu(p)u^\xi_\bp f_{\xi}(\bp) e^{-i p\cdot x}\nonumber \\
=&:\int\!\frac{d^3\bp}{(2\pi)^3}\varphi(\bp)e^{-i p\cdot x}
\end{align}
Hence the expectation value of the position in this scheme is
\be
\6\bx(t)\9=\int\!d^3\bx \bra\Psi\hat\Pi_\bx(t,\bx)\ket\Psi \bx=\int\!d^3 \bx\big|\Psi(t,\bx)\big|^2\bx,
\ee
while the expectation value of the momentum is simply
\be
\6\bp\9=\bra\Psi\hat\bp\ket\Psi.
\ee
Noting that
\be
\bnab_\bp e^{-i p\cdot x}=i(\bx-\bv t)e^{-ip\cdot x}, \qquad \bv=\bp /E_\bp,
\ee
and using Eq.~\eqref{onb}, we find
\be
\6\bx(0)\9=\sum_{\xi,\zeta}\int\!d\mu(p)f_\xi^*(\bp)u_\bp^{\xi\dag}\bnab_\bp\left(\frac{u_\bp^\zeta f_\zeta(\bp)}{2E_\bp}\right),
\ee
and  the expected relation
\begin{eqnarray}
\6\bx(t)\9 &=& \6\bx(0)\9+t\sum_{\xi}\!\int d\mu(p )f_\xi^*(\bp)f_\xi(\bp) \frac{\bp}{E_\bp} \nonumber \\
&=&\6\bx(0)\9+\6\bv \9 t.
\end{eqnarray}

For   future reference we note that (Appendix \ref{localap})
\begin{align}
&\6x_n^2(0)\9:=\int d^3\bx \,x_n^2 |\Psi(0,\bx)|^2\nonumber\\
&=\sum_{\xi,\zeta}\!\int\!\! d^3\bx\!\int\!\! d\mu(k)d\mu(p )f_\xi^*(\bk)f_\zeta(\bp)u_\bk^\xi{}\!^\dag u_\bp^\zeta \pad_{p_n}\!\pad_{k_n} \, e^{-i(\bk-\bp)\cdot\bx}\nonumber\\
&=\int\!\frac{d^3\bp}{(2\pi)^3}\big|\pad_{p_n}\varphi(\bp)\big|^2. \label{mom2x}
\end{align}

\vspace{1mm}

\subsection{Centre of energy POVM}
\vspace{-4mm}

Energy density broadly agrees with our intuition of ``where the particle is''. Indeed, the position variable that was based on it, \vspace{-2mm}
\be
\bq:=E^{-1}\int\!d^3\bx\,\bx T^{00},
\ee
and its quantum-mechanical analogues, were introduced in the early years of quantum mechanics [definition (c) of Pryce, \cite{pryce:48}; Appendix~\ref{twoop}].

The connection with energy density is most immediate for photons: when  electrons in a photodetector interact with the electric field of light, then a leading-order detection probability is proportional to the expectation value of the normal-ordered electric-field intensity
operator, and the latter is proportional to the energy density. There is no such  link with the models of  particle detectors. However, the picture is intuitively attractive and, as we find, corresponds very closely to the results of the previous section.

 We use the symmetrized normal-ordered energy density operator $::\!\!\hT_{00}\!\!::$   (see Appendix \ref{ap:pf}). To enforce the convexity of the trace formula, i.e., to maintain that the probability of a particular outcome for a weighted mixture of states is a weighted mixture of the corresponding probabilities,   normalisation should be performed at the level of operators \cite{t:14}. Hence the centre of energy POVM is constructed as
\be
\hPi_\bq(x)=::\hH^{-1/2}\hT_{00}\hH^{-1/2}::,
\ee
where $\hH$ is the field Hamiltonian. Its expectation value on a generic one-particle state $|\Psi\9$,
\begin{widetext}
\be
\bra\Psi\hat{\Pi}_\bq(t,\bx)\ket\Psi =\sum_{\xi,\zeta}\int\!d\mu (k) d\mu (p)T(\bp,\bk)f^*_\xi(\bk) u^{\xi\dag}_{\bk} u^{\zeta}_{\bp} f_\zeta(\bp) \,e^{i(q-p)\cdot x},
\ee

differs from its $\6\hPi_\bx\9$ counterpart by the presence of the factor
\be
T(\bp,\bk)=\frac{1}{2}\left(\sqrt{\frac{E_{\bp}}{E_{\bk}}}+\sqrt{\frac{E_{\bk}}{E_{\bp}}}\right).
\ee

 A lengthy, but straightforward calculation leads to
\be
\6\bq(0)\9:=\int d^3x \,\bx\bra\Psi\hat\Pi_\bq(0,\bx)\ket\Psi
 =\6\bx(0)\9+i\sum_{\xi,\zeta}d\mu(k)f^*_\xi(\bk) u^{\xi\dag}_{\bk} {u_\bk^\zeta f_\zeta(\bk)}\frac{1}{2\,E_\bk}
 \bnab_\bp T(\bp,\bk)\Big|
_{{\bp=\bk}}=\6\bx(0)\9,
\ee
because $\bnab_\bp T(\bp,\bk)\big|_{\bp=\bk}\equiv 0$.
Similarly,
\be
\6\bq(t)\9=\6\bx(t)\9=\6\bq(0)\9+\6\bv \9 t.
\ee

On the other hand, the expectations of squares of the position components are
\be
\6q^2_n(0)\9=\int\!d^{3}\bx\, x^2_n\bra\Psi\hat\Pi_\bq(0,\bx)\ket\Psi
=\6 x^2_n(0)\9+\sum_{\xi}d\mu(k)f^*_\xi(\bk)  { f_\xi(\bk)} \pad_{p_n}\pad_{q_n} T(\bp,\bk)\Big|
_{{\bp=\bk}}=\6x_n^2(0)\9-\left\langle\frac{p_n^2}{4E_\bp^4}\right\rangle,\label{rmx}
\ee
\end{widetext}
$n=1,2,3$. (See Appendix \ref{localap} for the details).
Hence the two POVMs for position do not coincide,
 in contradistinction to the coinciding Dirac operators $\tilde{\bx}$ and $\bq$, Eq.~\eqref{rxop}.

\subsection{Uncertainty relations}
Now we produce an estimate of the uncertainty relations. First we note that despite the non-commutativity of the Dirac operators $\tilde\bx$ and $\bq_\mathrm{Pr}$, the three statistical moments are obtained from the same probability measure and thus are simultaneously measurable. Since neither of the first moment operators
\be
\hat{x}_n^{(1)}:=\int d^3\bx\, x_n \hPi_\bx(x), \qquad \hat{q}_n^{(1)}:=\int d^3\bx\, x_n \hPi_\bq(x),
\ee
where $n=1,2,3$ is a   canonical conjugate of the momentum operator
\be
\hat{P}_n=\sum_\xi\int\!d^3\mu(p)p_n(\hat{b}^\dag_{p\xi}\hat{b}_{p\xi}+\hat{d}^\dag_{p\xi}\hat{d}_{p\xi}),
\ee
the Heisenberg uncertainty relations do not apply. The minima
\be
\min\Delta x_n \Delta p_n, \qquad \min\Delta q_n \Delta p_n,
\ee
where $(\Delta z)^2:=\6z^2\9-\6z\9^2$, should be found by minimization over all one-particle states.

For our purposes it is enough to consider a Gaussian momentum profile. More precisely, we choose the state $|\Psi\9$ such that $\chi_1(\bp)\equiv 1$ and
\be
g(\bp):=\frac{f(\bp)}{\sqrt{2E_\bp}}=N \exp\left(-\frac{(p_1-k)^2}{2\Gamma^2}-\frac{p_2^2+p_3^2}{2\Gamma^2}\right), \label{gstate}
\ee
where $k$ is a non-zero expectation of the momentum in the $x-$ direction, the width  $\Gamma \ll {m}$, and the normalization constant $N=(2\sqrt{\pi}/\Gamma)^{3/2}$.

The variance of the momentum is
\be
\Delta p_n^2=\Gamma^2/2,
\ee
while the calculation of the variance of $\bx$ is more cumbersome (Appendix C). In the leading order the quantity
 \be
\frac{p_1^2}{4E_\bp^4}=\frac{k^2}{4(m^2+k^2)}+\cO(\Gamma/m)^2,
 \ee
that determines the difference $\6 x_1^2(0)\9-\6  {q}_1^2(0)\9$ between the two variances is very small. However, it turns to be sufficient to reduce the product of uncertainties below $\half$.

   To obtain the analytical result we expand in the powers of $p_1-k$ and $p_2$, $p_3$ (Appendix C). The results are presented in Fig.~1.
\begin{figure}[htbp]
\includegraphics[width=0.49\textwidth]{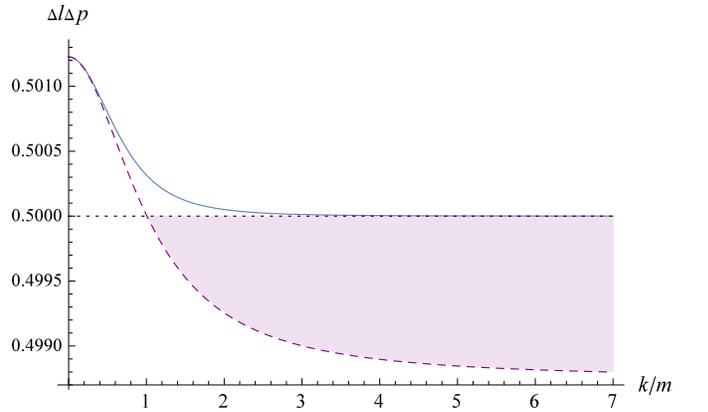}
\caption{Product of the standard deviations  in the units of $\hbar$: $\Delta x_1 \Delta p_1$ (thin blue line) and $\Delta q_1\Delta p_1$ (dashed purple line) compared with the Heisenberg bound $\half$ (dotted line). The curves correspond to the wave function \eqref{gstate} with $\Gamma=0.1m$.  }
\label{setup}
\end{figure}

The difference $\Delta  {q}_1\Delta p_1-\half$ becomes negative for $k\gtrsim m$.
\section{Dynamics}\label{sec:dyn}

We return to the model of a classical spin-$\half$ particle. It is applicable when the  uncertainties of Section~III are negligible relative to the scale of the action. The  resulting Eq.~\eqref{eomg} may be also used as the Heisenberg equation in the effective quantum mechanics of a particle in external fields.

The standard spin evolution equation   given  a trajectory in a   (constant) electromagnetic field  took its current form  in the work of  Bargmann, Michel, and Telegdi \cite{bmt}, but was essentially contained in the articles of Thomas \cite{thomas}, Frenkel \cite{f:26} and Tamm \cite{tamm:29}. Henceforth we refer to it as the TFT-BMT equation.

Dynamics of a  classical  particle of charge $e$ and mass $m$ in a given electromagnetic field $F^{\mu\nu}$ is determined by
\be
\dot{u}^\mu=w^\mu_e=\frac{e}{m}F^{\mu\nu}u_\nu.
\ee
 We allow for additional terms in acceleration that we collectively denote $w'$. These include influence of   other  forces \cite{reb:11}, such as the Newtonian gravity, and also take into account the spin  backreaction.
 While the latter naturally leads to the expansion in powers of spin, we restrict our discussion here only to the expression linear in $\sS$ (see, e.g., \cite{psk:00,s:00} for discussions of the higher-order spin terms).

Spin enters dynamics through its connection with the magnetic moment $\mu$. It is conveniently separated into the normal and the anomalous parts,
\be
\mu=\frac{g e}{2m}=\frac{e}{m} +\frac{(g-2)e}{2m}=\mu_0+\mu',
\ee
where the last expression   is suitable for neutral particles whose magnetic moment is wholly anomalous, $\mu=\mu'$.  We assume that the electric dipole moment is identically zero.

Following \cite{ll4}, a general form of the equation that is linear in external fields and the 4-spin is
\begin{eqnarray}
\frac{dS^\mu}{d\tau} &=&\alpha_0F^{\mu\nu}S_\nu+\alpha_2u^\mu  {F^{\nu\lambda}}u_\nu S_\lambda +\beta_1u^\mu w'^\nu S_\nu \nonumber\\
&+&\beta_2 \epsilon^{\mu\nu\lambda\rho}w'_\nu u_\lambda S_\rho, \label{spin1template}
\end{eqnarray}
where $\alpha_{0,2}$ and $\beta_{1,2}$ are constant coefficients. Using  the nonrelativistic  limit and conservation of the orthogonality relation $u^\mu S_\mu=0$, the first two terms result in the standard TFT-BMT equation
\be
\alpha_0=\mu, \qquad \alpha_2=-\mu'.
\ee
The third is analogous to the Thomas precession that takes into account $w'$
\be
\beta_1=-1.
\ee
The last term automatically satisfies conditions of Eq.~\eqref{polort}, so $\beta_2$ cannot be determined from kinematic considerations. However, in the rest frame  this term becomes $\beta_2 \boa'\!\times\bs$, leading to a parity-violating term in the Hamiltonian. While similar terms are expected to appear in the gravitational extension of the standard model \cite{sme,iw:16}, we are not going to consider them here. Moreover, we implicitly assumed that the laboratory reference frame is nonrotating. In practice, the effects of the Earth rotation, giving the Sagnac and the spin-rotation terms, should be taken into account.

To obtain the explicit three-dimensional form of the spin dynamics we use
the  equations of motion for   three-momentum $\bp$ and kinetic energy $E_\bp$,
\begin{subequations}
\begin{eqnarray}
&&\frac{d\bp }{dt}=e\bE+e\bv\times \bB+\frac{m^2}{E_\bp }\bw_{(0)}' \\
&&\frac{dE_\bp }{dt}=e\bv\cdot\bE+\frac{m^2}{E_\bp }\bv\cdot\bw_{(0)}', \label{eom0}
\end{eqnarray}
\end{subequations}
that take into account the additional acceleration.  {(The total energy $E$ of a spinless charged particle in an external electromagnetic field  satisfies $E=E_\bp+e A^0$, where $A^0$ is the scalar potential)}. The subscript $(0)$ indicates the order of the expansion in powers of spin, and we omitted it from the momentum, velocity and energy variables to reduce the clutter. The acceleration  $\bw$ is the spatial part of the four-acceleration, and $\bw'=\bw-\bw_e$. In the linear  approximation to Eq.~\eqref{spin1template} the above equations do not include spin.   As a result, we  obtain the standard three-dimensional form of the spin-precession equation,
\be
\frac{d\bS_{\mathrm{W}}}{dt}=\bG\times\bS_{\mathrm{W}}
\ee
where now
\begin{align}
\bG=&-\left(\frac{m}{E_\bp}\mu_0+\mu'\right)\bB+\frac{E_\bp}{E_\bp+m}\mu'(\bv\cdot\bB)\bv \nonumber \\
&-\left(\frac{m}{E_\bp+m}\mu_0+\mu'\right)(\bE\times\bv) \nonumber \\
&-\frac{E_\bp^2}{m(E_\bp+m)}(\bv\times\boa'_{(0)}), \label{gterm}
\end{align}
$\boa:=d\bv/dt$, and $\boa'_{(0)}$ is the three-dimensional counterpart of $\bw'_{(0)}$. Incidentally, using $\bS_{\mathrm{W}}$ instead of the numerically equal rest frame spin $\bs$ gives a consistent geometric meaning to the equation: all the quantities on the right hand side are defined in the same (laboratory) frame.

Lagrangian and Hamiltonian formulations of the dynamics of spinning particles can use  either a three- or four-dimensional approach. While questions of covariance of this dynamics, derivations of the TFT-BMT equation from an action principle, and quantization of the effective classical theory as to make a better connection with the fundamental Dirac-equation based field-theoretical analysis or the non-Abelian Berry phase  are more advantageously discussed in the four-dimensional calculations, we utilize a simpler three-dimensional form.

 We model our construction on the Derbenev-Kondratenko (DK)  Hamiltonian  \cite{dk:73}. It is built as a minimal combination of the (linear) relativistic spin precession and the Lorentz force.  It is also equivalent to the description proposed by Frenkel \cite{f:26} and is non-manifestly Poincar\'{e}-covariant. An alternative Hamiltonian  \cite{s:08} is based on the Foldy-Wouthuysen transformation \cite{ll4, tha:92}. Comparisons with the semiclassical expansion of the Dirac equation \cite{wbk:15} showed  that the latter Hamiltonian gives a better approximation starting from moderate energies $E_\bp\sim m$. However, the difference between the predicted forces is highly oscillatory and becomes significant on the scales of a fraction of the  atomic unit of length. Since we are interested in the effects on a much coarser scale (Section~V), we follow the DK construction.

The Wigner spin provides us with the natural canonical variables, since unlike many alternative spin operators, the triple $\hat{S}_\mathrm{W}^k$ satisfies the angular momentum commutation relations, implying the standard angular momentum Poisson brackets
\be
[S_{\mathrm{W}}^i,S_{\mathrm{W}}^j]_\mathrm{PB}=\epsilon^{ij}_{~~k}S_{\mathrm{W}}^k.
\ee
The spin potential energy is as $U_S=\bG\cdot\bS_{\mathrm{W}}$. The simplest way to obtain the equations of motion is to use the Rauthian function, that is  the Lagrangian in positions and velocities and the Hamiltonian in spin variables, the DK Rauthian
\begin{eqnarray}
R_\mathrm{DK} &=&-m\sqrt{1-\bv^2}+e\bA\cdot\bv-eA^0 \nonumber \\
&-&V(\bx,\bv) -\bG\cdot\bS_{\mathrm{W}}, \label{rdk}
\end{eqnarray}
where $V(\bx,\bv)$ generates  $\bw'$   in Eq.~\eqref{eom0}. As a result, the equation of motion is
 \be
\frac{d\bp}{dt}=e\bE+e\bv\times \bB+\frac{m^2}{E_\bp}\bw'_{(0)}+\frac{d}{dt}\frac{\pad\bG}{\pad \bv}\cdot\bS_{\mathrm{W}}-\frac{\pad\bG}{\pad \bx}\cdot\bS_{\mathrm{W}} \label{eomg}
\ee
The term $V(\bx,\bv)$ may describe a free fall of a nonrelativistic particle in an Earth-bound frame. It contributes to the spin precession via
\be
\bG_g\simeq-\frac{1}{2c^2}\bv\times\bg,
\ee
where $\bg$ is the free-fall acceleration and we restored $c$. This term can be obtained from the analysis of the Dirac equation on a curved background \cite{g-sg}.

\section{Discussion}

The Schr\"{o}dinger--Pauli equation \cite{ll4} describes spinning  electrons in  nonrelativistic wave mechanics.  A triple of operators --- halves of the Pauli matrices--- are both dynamical variables (that act on the two-component wave function as part of the Hamiltonian) and observables. Pauli matrices set the standard for the expected properties of a spin observable, and their role as generators of the symmetry group was taken on in relativistic field theories. The equation itself is obtained by taking the nonrelativistic limit of the Dirac equation and identifying the large components of the Dirac spinors as a two-component wave functions. Similarly, an explicit role of spin  (as expressed by the Pauli matrices) in nucleon interactions is obtained through the effective field theory Lagrangians for the low energy QCD \cite{eft-n}

Covariance properties under the relevant symmetry group, often as an abstraction of the properties of classical variables, play a role in construction of quantum observables \cite{peres:95,bu:99, cov-povm}. Additional requirements, such as satisfying particular commutation relations or properties of a nonrelativisitc limit, can be added. Three reasonable requirements single out the Wigner spin \cite{bogo:90,t:03}.

Effective dynamics  is written in terms of the emergent (or guessed) variables. The criteria are the ease of analysis and quality of the resulting approximation. As we have seen in Section~\ref{sec:dyn} the dynamics of classical spinning particles can be described either in terms of the 4-vector of spin or Wigner spin. Moreover, since there are unambiguous relationships between different versions of spin \cite{pryce:48,7op,cz:97} the search for the ``best'' spin observable  e{becomes} a choice of the most convenient representation of data for a particular purpose.

Dynamics, whether of a high-energy scattering problem or of a particle in given external fields, also can be analyzed using different versions of spin. Again, depending on the problem, it may be more convenient to use either helicity, or four-polarisation, or the rest frame Wigner spin.

The Stern-Gerlach experiment \cite{peres:95,cov-povm} is the standard theoretical description of spin observation. To avoid  complications due to the Lorentz force \cite{accel1} we consider  motion of neutrons in a classical magnetic field \cite{sg-n}. Equations of motion are most conveniently obtained by using $\bS_{\mathrm{W}}$ that have standard commutations/Poisson brackets relations. Eq.~\eqref{gterm} becomes
\be
\bG=-\mu\bB+\frac{E_\bp}{E_\bp+m}\mu (\bv\cdot\bB)\bv\approx -\mu\bB+\half\mu \bv (\bv\cdot\bB), \label{gneutron}
\ee
where we kept only the leading relativistic corrections. Nevertheless, once they are written,  by virtue of Eq.~\eqref{eomg} any version of spin can be used in the analysis. The real question is  what are the actual predictions in a realistically modelled magnetic field \cite{cov-povm}, particularly noting that a careful analysis of a nonrelativistic scenario indicated that the results are much less sharp than those presented in the cartoon depictions of the experiment.

 Several directions follow from this work. A significance of the invariant $\6\sS\9^2$ should be clarified. Analysis of  \cite{dpm:14b} points at the Pryce's (d)-type position, essentially a Lorentz-transformed centre of mass, as preferred set of operators. We plan to extend the POVM formalism to this case. Analysis of the relativistic version of the Stern-Gerlach experiment \cite{sv:12a,ptw:13} produced some qualitatively new features. Using Eq.~\eqref{gneutron} in the the equation of motion \eqref{eomg}, together with a realistic profile of the magnetic field will complete the idealized picture of the experiment.

 {The \textit{Zitterbewegung} effect of velocity oscillations  around the average value $\6\bp H^{-1}\9$ results from superposition of positive- and negative-energy solutions \cite{tha:92}. It is a mathematical artefact for free Dirac particles, as well as particles in not-too-strong electromagnetic fields. The position POVMs $\hPi_\bx$ and $\hPi_\bq$, as well as the corresponding Dirac equation operators (Appendix~\ref{twoop}), separate the electron and positron states and eliminate this effect.  A controversial result of a transverse force exerted by an external electric field is closely related to this phenomenon \cite{trfo}. We expect that unlike its condensed matter counterpart \cite{zb-c} this effect will disappear for properly localized wave packets and will investigate the localization in external fields.   }

 A more realistic description that takes into account a finite extent of the wave packets (Section~\ref{local3}) will produce a relativistic counterpart  of the analysis in \cite{cov-povm}. Finally, a natural next step is to extend the analysis of Section~\ref{local3} to curved space-times, connecting to the results derived from the Dirac equation.

\acknowledgments
This work was supported in part by the Brazilian funding agencies CNPq (Grants No. 401230/2014-7, 445516/2014-3 and 305086/2013-8), CAPES, and the National Institute for Quantum Information (INCT-IQ). DRT thanks Technion --- Israel Institute of Technology for hospitality and Technion Australia for financial support.  We thank  Alexei Deriglazov, Juan L\'{e}on, Robert Mann, Daniel Martinez, Pablo Saldanha, Amos Ori, and Vlatko Vedral for comments and discussions.

\appendix

\section{Frames and Lorentz transformation} \label{ap:lorentz}
\subsection{Conventions} \label{conven}


The totally antisymmetric four-dimensional symbol is defined by $\epsilon_{0123}:=+1$, i.e. $\epsilon^{0123}:=-1$.
Purely spatial three-dimensional symbols $\epsilon_{ijk}=\epsilon^{ijk}=\epsilon^{ij}_{~~k}$ are always defined with respect to the Euclidean signature $+++$, i.e $\epsilon^1_{~23}=+1$. %

\subsection{Lorentz transformations and kinematics}
The lab frame $\cS_L$ and a frame $\cS'$ are related by a Lorentz transformation $\Lambda(\bv,R)$, where $\bv$ as a velocity of (the origin of) $\cS'$ relative to $\cS_L$, and $R$ represents the three rotation parameters either as a 3-D matrix $R$ or in any other form.  The vector components in the two frames are related by  $x'^\mu=\Lambda^\mu_{~\nu}x^\nu$ (a passive transformation). The standard reference momentum for massive particles is $p_s^{\,\mu}=(m,0,0,0)$. The standard Lorentz transformation $L_p$ takes it to $p$  {(an active transformation)},  hence
$L_p=\Lambda(-\bv,0)=\Lambda(\bv,0)^{-1}$ is
 \be
L_p=
\begin{pmatrix}
 u^0 & u^1 & u^2 & u^3 \\[2pt]
 u^1 & 1+\dfrac{(u^1)^2}{1+u^0} & \dfrac{u^1 u^2}{1+u^0} & \dfrac{u^1 u^3}{1+u^0} \\[8pt]
 u^2 &  \dfrac{u^1 u^2}{1+u^0} & 1+\dfrac{(u^2)^2}{1+u^0} & \dfrac{u^1 u^3}{1+u^0} \\[8pt]
 u^3 & \dfrac{u^1 u^3}{1+u^0} & \dfrac{u^2 u^3}{1+u^0} & 1+\dfrac{(u^3)^2}{1+u^0} &
\end{pmatrix}
\ee

Since the 4-velocity is $u=(u^0,\bu)=\gamma(1,\bv)$, and
\be
\dot\gamma=\gamma\frac{d}{dt}(1-v^2)^{-1/2}=\gamma^4\bv\cdot\frac{d\bv}{dt}=\gamma^4\bv\cdot\boa,
\ee
the 4-acceleration is given explicitly in terms of the three-dimensional quantities as
\begin{align}
w:=&\dot u=(w^0,\bw)=\dot\gamma(1,\bv)+\gamma^2(0,\bv\,') \nonumber \\
=&(\gamma^4\bv\cdot\boa,\gamma^4(\bv\cdot\boa) \bv+\gamma^2\boa).
\end{align}

\section{Dirac equation}\label{ap:dirac}
\subsection{Dirac matrices and spinors}\label{spinor}
We use the standard representation of the Dirac matrices,
\be
\beta=\gamma^0, \qquad \alpha^k=\gamma^0\gamma^k,
\ee
and we set
\be
\gamma^5=-i\gamma^0\gamma^1\gamma^2\gamma^3.
\ee
The standard (Dirac) spin operator is
\be
 \bm{\Sigma}=-\balp \gamma^5=-\frac{i}{2}\balp\times\balp.
\ee

Dirac Hamiltonian is given by
\be
H=\balp\cdot\bp+\beta m=-i\balp\cdot\bnab+\beta m,
\ee
and the angular momentum is given by the version of Eq.~(1) as}
\be
\bJ=\bx \, \times \, \bp+\half\bmsig.
\ee

We use the symmetric form of the Lagrangian density,
\begin{eqnarray}
\cL &=&\bar{\psi}(i\gamma^{\mu}\partial_{\mu}-m)\psi-\half\big(i\pad^\mu\bar\psi)\gamma_\mu\psi \nonumber\\
&=&\half\bar\psi i\gamma^\mu\overleftrightarrow\pad\!_\mu\psi-m\bar\psi\psi,
\end{eqnarray}
that results in the symmetric energy-momentum tensor. In particular,
\be
T_{00}=\half i\psi^\dag\overleftrightarrow\pad\!_0\psi-\half i\bar\psi\gamma^\mu\overleftrightarrow\pad\!_\mu\psi+m\bar\psi\psi.
\ee

The basis of positive- and negative energy solutions of the Dirac equation, $\psi_p(x)=u_\bp^\alpha e^{-ip\cdot x}$ and $\psi_{-p}(x)=v_\bp^\alpha e^{-ip\cdot x}$, respectively, are given in the standard representation as
\begin{subequations}
\begin{align}
u_\bp^\alpha&=\frac{1}{\sqrt{(E_\bp+m)}}\left(\begin{matrix}
(E_\bp+m)\chi_\alpha\\ \bp\cdot\bm{\sigma}\chi_\alpha
\end{matrix}\right), \\
 v_\bp^\alpha&=\frac{1}{\sqrt{(E_\bp+m)}}\left(\begin{matrix}
\bp\cdot\bm{\sigma}\chi_\alpha \\(E_\bp+m)\chi_\alpha \end{matrix}\right),
\end{align}
\end{subequations}
where the rest-frame spin is given by the two-dimensional spinors
\be
\chi_{1\!/2}=\left(\begin{matrix}1\\0\end{matrix}\right), \qquad \chi_{-1\!/2}=\left(\begin{matrix}0\\1\end{matrix}\right).
\ee
Their orthogonality and complexness relations are normalized as
\begin{subequations}
\begin{align}
&\bar{u}_\bp^\alpha u_\bp^\beta=2m\delta^{\alpha\beta}, \qquad \bar{v}_\bp^\alpha v_\bp^\beta=-2m\delta^{\alpha\beta} \\
&u_\bp^\alpha{}^\dag u_\bp^\beta=v_\bp^\alpha{}^\dag v_\bp^\beta=2E_\bp\delta^{\alpha\beta} \label{onb}\\
&\sum_\alpha u_\bp^\alpha u_\bp^\alpha{}^\dag+v_{-\bp}^\alpha{} v_{-\bp}^\alpha{\!\!}^\dag=2E_\bp I_{4\times 4},
\end{align}
\end{subequations}
where $I_{4\times 4}$ is the four-dimensional identity matrix.

\subsection{Position operators for the Dirac equation}\label{twoop}
 {The multiplicative  position operator $\bx$ that corresponds via Eq.~(1) to the Dirac spin $\bS_{\mathrm{D}}$ does not preserve the positive- and negative-energy subspaces.}
Replacing it with the one that leaves the positive and negative energy subspaces separately invariant gives \cite{tha:92}
\be
\tilde{\bx}=P_+\bx P_++P_-\bx P_-=\bx+2iH^{-1}\bF,
\ee
where $P_\pm$ are the projections on the positive (negative) energy spaces, and
\be
\bF:=\balp-\bp H^{-1}.
\ee

On the other hand, Pryce's position operator (c) \cite{pryce:48} is a non-commutative generalization of the centre of energy $\bq=\bN/E_\bp$, where
\be
\bN=\half(\bx H-H\bx)=\bx H-\half i\balp.
\ee
Then
\be
\bq=\bx+\frac{1}{2E_\bp^2}\left(\bp\times\bm{\Sigma}+im\beta\balp\right).
\ee
By using the commutation and anticommutation relations of $H$, $\bF$, and $\balp$, as well as the identity $H^{-1}=H^2/E$, we observe that
\be
\tilde\bx=\bq.
\ee

{Applying the definition of the position operator (c) in a centre of mass frame (i.e., the one with the zero momentum), and Lorentz-transforming to an arbitrary frame defines Pryce's position operator (d) \cite{pryce:48,dpm:14b}. If the goal is to obtain a triple of pairwise commuting operators, then the  average of the positions (c) and (d), weighted by the energy and the rest mass respectively, acheaves it. Namely,
\be
\tilde\bq=(E\bq+m\mathbf{X})/(m+E)
\ee
gives Pryce's position variable (e) in the classical case, where $\mathbf{X}$ is Pryce's position (d). Its quantum version is the Newton-Wigner position operator \cite{nw:49}.

\subsection{Spin operators for the Dirac equation}\label{twospin}
The spin operator that is associated with the position operator $\bq=\tilde\bx$,
\begin{align}
\tilde\bmsig&=P_+\bmsig P_++P_-\bmsig P_-=-\frac{i}{4}\bF\times\bF\nonumber\\
&=\frac{1}{2E_\bp}\big(m^2\bmsig-im\beta\balp\times\bp+(\bmsig\cdot\bp)\bp\big)\nonumber\\
&=\bJ-\bq\times\bp\equiv \bS_\mathrm{Cz},
\end{align}
is equivalent to the operator that was introduced by Czachor \cite{cz:97,7op} using different considerations.}

{The position operator $\mathbf{X}$ corresponds via Eq.~(1) to
\be
\bS_\mathrm{F}:=\bJ-\bX\times\bp=\half\big(\bmsig-i\beta\balp\times\bp\big),
\ee
that can be traced to Frenkel, Ref.~\cite{f:26}.

{Finally, the Newton-Wigner position operator $\tilde\bq$ corresponds to the Wigner spin $\bS_{\mathrm{W}}$ \cite{7op}. In terms of Dirac matrices it can be written, e.g., as
\be
\bS_{\mathrm{W}}=\frac{1}{2E}\big(m\bmsig-i\beta\balp\times\bp\big)+\frac{(\bp\cdot\bmsig)\bp}{2E(E+m)}.
\ee
Using Eqs.~\eqref{stsp} and \eqref{stsp} the direct evaluation establishes that, indeed,
\be
\6\hbS_{\mathrm{W}}\9=\half\chi^\dag\bm\sigma\chi=\frac{1}{2E_\bp}u^\dag_\chi(\bp){\bm\bS_{\mathrm{W}}}u_\chi(\bp)
\ee

{We summarize the classifications of some of the Dirac equation operators in Table~I.}

\appendix
\Alph{section}
\setcounter{section}{2}
\section{Conventions for particles and fields}\label{ap:pf}

We define the basis states of particles and antiparticles as
\be
|p,\sigma\9=\hb_{p\sigma}^\dag|0\9, \qquad |q,\sigma\9_\mathrm{a}=\hd_{p,-\sigma}^\dag|0\9,
\ee
respectively, and creation and annihilation operators satisfy the anticommutation relations,
\be
[\hb_{p\xi},\hb_{q\zeta}^\dag]_+=[\hd_{p\xi},\hd_{q\zeta}^\dag]_+=(2\pi)^3(2E_\bp)\delta^{(3)}(\bp-\bk)\delta_{\xi\zeta} \label{rxop}
\ee

The explicit construction of spin states  begins with picking a reference four-momentum $p_s$. The Wigner spin  and other spin operators are defined to  coincide with the nonrelativistic spin  in particle's rest frame.

The one-particle basis states are defined by
\be
|p,\sigma\9\:=\hat U[L_p]|p_s,\sigma\9,
\qquad\hat{S}_\mathrm{W}^3|p,\sigma\9=\sigma|p,\sigma\9. \label{basis}
\ee

Using the group representation property and Eqs.~(\ref{basis}) the transformation  is written as
\be
\hat U(\Lambda)=\hat U[L_{\Lambda p}]\hat U[L^{-1}_{\Lambda p}\Lambda L_p]\hat U[L^{-1}_p],
\ee
where the element of the Lorentz group
\be
\cW(\Lambda,p):= L^{-1}_{\Lambda p} \Lambda L_p, \label{wigw}
\ee
leaves $p_s$ invariant, i.e.  belongs to the stability subgroup (or Wigner little group) of $k_R$. Finally,
\be
\hat U(\Lambda)|p,\sigma\9=\sum_\xi D_{\xi\sigma}[\cW(\Lambda,p)]|\Lambda
p,\xi\9,
\ee
where $D_{\xi\sigma}$ are the matrix elements of the representation of  $\cW(\Lambda,p)$. For our choice of $p_s$ the little group consists of rotations, and for  spin-$\half$ any $2\times2$ unitary matrix can be written as ${D}=\exp(-i\omega \hat{\mathbf n}\cdot \mbox{\boldmath $\sigma$})$, where $\omega$ is a rotation angle and $\hat{\mathbf n}$ is a rotation axis that corresponds to $\cW(\Lambda,p)$.

With these conventions the  Dirac field is written as
\begin{eqnarray}
\hat\psi &=& \sum_{\xi}\int\!d\mu(p)\big(\hb_{p\xi}u_\bp^\xi e^{-ip\cdot x}+\hd_{p\xi}^\dag v_\bp^\xi e^{i p\cdot x}\big) \nonumber \\
&=:& \hpsi^{(+)}(x)+\hpsi^{(-)}(x).
\end{eqnarray}
The energy density is given by
\begin{align}
&::\hat{T}_{00}(t,\bx)::= \half\sum_{\xi,\zeta}\int\!d\mu(k)d\mu(p) (E_\bp+E_\bk)\times\nonumber \\
&\big(u^{\xi\dag}_\bk u^{\zeta}_\bp\hat{b}^\dag_{q\,\xi}\hat{b}_{p\,\zeta}e^{i(q-p)\cdot x}+v^{\xi\dag}_\bk v^\zeta_\bp \hat{d}^\dag_{p\,\zeta}\hat{d}_{q\,\xi} e^{-i(q-p)\cdot x}\big) \nonumber \\
&+\ldots
\end{align}
where $\ldots$ stand for terms whose expectations on one-particle states vanish. The field Hamiltonian is
\be
\hat{H}=\sum_\xi\int d\mu(p){{E_\bp}}(\hat{b}^\dag_{p\xi}\hat{b}_{p\xi}+\hat{d}^\dag_{p\xi}\hat{d}_{p\xi}),
\ee
and the restriction of its inverse square root to the one-particle states is
\be
\hat{H}^{-1/2}=\sum_{\xi}\int d\mu(p)\frac{1}{\sqrt{E_\bp}}(\hat{b}^\dag_{p\xi}\hat{b}_{p\xi}+\hat{d}^\dag_{p\xi}\hat{d}_{p\xi}).
\ee

\section{Details of the localization calculations}\label{localap}

First we provide the explicit form of the terms in  $\6 {q}_n^2(0)\9$, Eq.~\eqref{rmx}. We re-write it as
\begin{align}
\6 {q}_n^2(0)\9=\int\!d^3\bx\!\int \frac{d^3\bp}{(2\pi)^3} \frac{d^3\bk}{(2\pi)^3}e^{-i(\bk-\bp)\cdot \bx} \nonumber \\
\times \pad_{p_n}\!\pad_{k_n}\Big(\varphi(\bp)^\dag\varphi(\bk)T(\bp,\bk)\Big),
\end{align}
Noting that $T(\bp,\bp)\equiv 1$ and $\pad_{p_n}T(\bp,\bk)\big|_{\bk=\bp}\equiv 0$, we obtain
\begin{align}
\6 {q}_n^2(0)\9=&\int \frac{d^3\bp}{(2\pi)^3}|\pad_{p_n}\varphi(\bp)|^2 \nonumber\\
+&\int\!\frac{d^3\bp}{(2\pi)^3}|\varphi(\bp)|^2\pad_{p_n}\!\pad_{q_n}T(\bp,\bk)\Big|_{\bp=\bk}.
\end{align}
Taking into account that
\be
\pad_{p_n}\!\pad_{k_n}T(\bp,\bk)\Big|_{\bp=\bk}=-\frac{p_n^2}{4 E_\bp^2},
\ee
and Eq.~\eqref{onb} we obtain Eq.~\eqref{rmx}.

For a state  with $\chi_1(\bp)\equiv 1$ and, e.g., the momentum profile of Eq.~\eqref{gstate},  Eq.~\eqref{mom2x} leads to
\begin{widetext}
\begin{align}
\6x_n^2(0)\9&=\int \frac{d^3\bp}{(2\pi)^3}|\pad_{p_n}\varphi(\bp)|^2  \nonumber \\
&=\int\!\! \frac{d^3\bp}{(2\pi)^3}\left( \left|\frac{\pad}{\pad p_n}\frac{f(\bp)}{2E_\bp} \right|^22E_\bp+\left|\frac{f(\bk)}{2E_\bk} \right|^2 \left|\frac{\pad u_\bp^1}{\pad p_n}\right|^2 +
\left(\frac{\pad}{\pad p_n}\frac{f^*(\bp)}{2E_\bp}\right)\frac{f(\bp)}{2E_\bp}  \frac{u_\bp^1{}\!^\dag\pad u_\bp^1}{\pad p_n}
+\frac{f^*(\bp)}{2E_\bp}\left(\frac{\pad}{\pad p_n}\frac{f(\bp)}{2E_\bp}\right) \frac{\pad u_\bp^1{}\!^\dag}{\pad p_n} u_\bp^1\right) \nonumber  \\
&=\int\!\! \frac{d^3\bp}{(2\pi)^3}\left( \left|\frac{\pad}{\pad p_n}\frac{g(\bp)}{\sqrt{2E_\bp}} \right|^22E_\bp+\left|\frac{g(\bk)}{\sqrt{2E_\bk}} \right|^2 \left|\frac{\pad u_\bp^1}{\pad p_n}\right|^2 + 2\mathrm{Re}\left[
\left(\frac{\pad}{\pad p_n}\frac{g^*(\bp)}{\sqrt{2E_\bp}}\right)\frac{g(\bp)}{\sqrt{2E_\bp}}  \frac{u_\bp^1{}\!^\dag\pad u_\bp^1}{\pad p_n}\right]
\right).
\end{align}
\end{widetext}
The leading term expansion gives
\be
\6x_n^2(0)\9=\frac{1}{4}\frac{1}{\Gamma^2}+\frac{m^2}{E_k^2}+\cO(\Gamma^2),
\ee
where $E_k=\sqrt{k^2+m^2}$ and the $\cO(\Gamma^2)$ term scales as $\Gamma^2/m^4$ in the nonrelativistic limit and as $\Gamma^2/k^4$ in the ultrarelativistic case.

\end{document}